# Extending the Hegselmann-Krause Model of Opinion Dynamics to include AI Oracles

Allen G. Rodrigo
School of Biological Sciences
The University of Auckland
Email: a.rodrigo@auckland.ac.nz

Abstract: The Hegselmann-Krause (HK) model of opinion dynamics describes how opinions held by individuals in a community change over time in response to the opinions of others and their access to the true value, $\tau$, to which these opinions relate. Here, I extend the simple HK model to incorporate an Artificially Intelligent (AI) Oracle, that averages the opinions of members of the community. Agent-based simulations show that (1) if individuals only have access to the Oracle (and not $\tau$), and incorporate the Oracle's opinion as they update their opinions, then all opinions will converge on a common value; (2) in contrast, if all individuals also have access to $\tau$, then all opinions will ultimately converge to $\tau$, but the presence of an Oracle may delay the time to convergence; (3) if only some individuals have access to $\tau$, opinions may not converge to $\tau$, but under certain conditions, universal access to the Oracle will guarantee convergence to $\tau$; and (4) whether or not the Oracle only accesses the opinions of individuals who have access to $\tau$, or whether it access the opinions of everyone in the community, makes no marked difference to the extent to which the average opinion differs from $\tau$.

**Introduction**

The Hegselmann-Krause (HK) model (Hegselmann and Krause 2002, Hegselmann and Krause 2006, Hegselmann and Krause 2009) has been widely used to study how individuals might alter their opinions in the light of the opinions of others, particularly those who may share similar views. The HK model, in its simplest incarnation is a relatively straightforward time-difference equation where, at each time-step, the $i^{th}$ individual updates a real-valued opinion they hold, $x_i(t)$ (bounded between 0 and 1, inclusive), based on the (possibly, weighted) average of the opinions of other individuals that are no more different from $x_i(t)$ than some threshold, $\varepsilon$. The value $\varepsilon$ establishes a bounded-confidence envelope (BCE) around each individual; thus, the HK model is one of a class of bounded-confidence models.

There are two typical outcomes with the simple HK model: time proceeds until either all individuals share the same opinion (i.e., a consensus is reached), or groups of individuals share common opinions not shared with other groups (i.e., polarization). Which of these outcomes arise typically depends on the size of the BCE.

In 2006, Hegselmann and Krause introduced a term in the model that expresses an individual's access to the Truth, $T$, i.e., the real value that each individual has an opinion about. In the simplest case, $T$ is univariate although there has been some investigations of opinion dynamics when $T$ is multivariate (Riegler and Douven 2009, Etesami, Basar et al. 2013). What might the value of $T$ represent? Concrete examples might include the total number of named mammal species (Burgin, Colella et al. 2018), or the increased risk of developing lung cancer in individuals who vape and smoke compared to individuals who only smoke (Bittoni, Carbone et al. 2024). A more mundane, but rather quaint and charming example of what the value of $T$ might be comes to us courtesy of Francis Galton (Galton 1907). In 1906, at the West of England Fat Stock and Poultry Exhibition at Plymouth, about 800 individuals entered a competition to guess the weight of an ox. Galton was given access to the guesses of the ox's weight, and calculated their average value to be 1207 pounds. The real weight of the ox was 1198 pounds, a mere 0.8% difference (Galton 1907).

Galton noted that "The competitors included butchers and farmers, some of whom were highly expert in judging the weight of cattle; others were probably guided by such information as they might pick up, and by their own fancies. The average competitor was probably as well fitted for making a just estimate of the dressed weight of the ox, as an average voter is of judging the merits of most political issues on which he votes, and the variety among the voters to judge justly was probably much the same in either case" (Galton 1907). In saying as much, Galton anticipates the HK model, whereby an individual might access the Truth through some reliable experimental, statistical or heuristic method, or through accumulated experience. The model is, of course, agnostic about process, and quantifies access to $T$ by a weight or coefficient within the equation that does not distinguish amongst the different processes that individuals might use to access $T$. Interestingly, Hegselmann and Krause (Hegselmann and Krause 2006) demonstrated mathematically that when all individuals in a community have access to $T$, then ultimately, all opinions will converge to $T$. In the same paper, Hegselmann and Krause also showed that if only some individuals have access to $T$ – what they referred to as a "cognitive division of labor" – then polarization can occur depending on the weight placed on $T$ by those who have access to it.

Since these early developments of the HK model, it has been used extensively in a wide range of studies spanning a wide range of disciplines. A survey of recent papers reveal the application of the HK model in diverse offerings, including: strategies associated with the communication and transmission of health information (Wei, Qin et al. 2022, Yin, Tang et al. 2024), the dynamics of social networks (Li, Li et al. 2022), the consequences of suppressing free expression of opinions (Peng, Zhao et al. 2025), policy-making (Lammers, Pattyn et al. 2024), the consequences of persuasion (Xu, Xiao et al. 2023), the evolution of inductive reasoning (Douven 2023), comparisons of authoritarianism and democracy in the face of disease outbreaks (Biondo, Brosio et al. 2022), the influence of opinion leaders (Chen et al 2020), and so on.

Additionally, the relatively simple model first developed by Hegselmann and Krause (Hegselmann and Krause 2002) has lent itself to quite sophisticated mathematical analyses with extensions that include network and graph theory (Li 2024), game theory (Jond 2024), multidimensional opinion vectors (Etesami, Basar et al. 2013, Cheng, Chen et al. 2025), and the addition of noise to opinion dynamical systems (Su, Chen et al. 2017, Chen, Nikolaev et al. 2025).

In this paper, I return to the relatively simple form of the original HK model, and I extend it by incorporating an accessible Artificial Intelligence (AI), specifically, an AI that functions as an Oracle (Armstrong, Sandberg et al. 2012). Armstrong et al define an Oracle as "an AI that does not act in the world except by answering questions". More recently, Messeri and Crockett (Messeri and Crockett 2024) develop a taxonomy of the scientific community's "visions" of what AI might become, and what purposes it might serve. Of the four categories that Messeri and Crockett identify, one is an Oracle, something "that can digest and communicate scientific knowledge [and] promises to solve an important problem: the deluge of published material 'threatening to exceed the cognitive limits of human processing capacities …".

Therefore, an Oracle has three important features: first, it provides answers to questions (including, for our purposes, the question, "What is the value of $T$"?); second, it does so by collating and synthesizing all relevant information; and third, by implication, the Oracle has no primary access to $T$, only secondary access via the community from which it obtains its information.

Given the extraordinary interest in Large Language Models (LLMs) and Generative AI (gAI) over the last few years, it is timely and important to examine how AI, particularly AIs that function as Oracles, can influence the opinions of those in the community. One might argue, of course, that Oracles *sensu* Armstrong et al (Armstrong, Sandberg et al. 2012) did not begin with LLMs and gAI; after all, the first search engines that appeared in the 1990s (e.g., Metacrawler, Yahoo!, Google, LiveSearch, Ask.com, etc.) answered users' questions by directing users to relevant information from across the nascent World Wide Web. However, current AI tools, including gAIs, introduce additional opportunities and challenges beyond those encountered with search engines. In particular, gAIs can be trained to answer to questions directly, based on the corpora of digital resources from an entire community. Alternatively, they may be trained only on resources that are domain-specific, for example, scientifically-relevant databases, or open-access publications and code. In fact, domain-specific training is often seen as a solution to the propensity for LLM-based gAIs to "hallucinate", that is, to sometimes (often?) provide false or misleading information (Cinquin 2025, Liu, Yin et al. 2025).

In this paper, I have two overarching questions: first, under what conditions will the presence of Oracles lead to polarisation or consensus; and second, under what conditions will the presence of Oracles lead the community to converge to, or diverge from, the Truth. In the sections that follow, I first develop the simplest form that an Oracle might take in the HK model, one where the Oracle synthesizes all prevailing opinions in the community of individuals by taking an unweighted average of all opinions in the community. I show, by simulation, how the dynamics of opinion change with and without the Oracle, both when the Truth, $T$, is accessible and when it is inaccessible to members of the community. (Note: in this manuscript I use $T$ to refer to Truth as a named variable, and its value by $\tau$).

I next modify the model to allow only some individuals to have access to the Oracle, and only some individuals to have access to $T$; additionally, I allow the Oracle to construct its opinion based on a subset of opinions, rather from the entire community of opinions. Thus, an Oracle might only take the average of opinions of those individuals who have access to $T$, or alternatively, only those individuals who do not have access to $T$.

I have avoided mathematical solutions and analyses of our model and their variants; instead, I have opted to implement these as computational Agent Based Models (ABMs). The results I show below are those obtained by simulations; thus, they provide a qualitative picture of how Oracles may function in a community of opinionated individuals.

Prior to conducting the simulations, I had clear hypotheses or conjectures of what our results would be. In hindsight, it is perhaps unsurprising that these were not always borne out. It is useful, therefore, to list what these conjectures are before proceeding to the results – effectively, the simulations are tests of these conjectures:

**Conjecture 1**. In the absence of access to $T$, communities will converge to a common opinion if all individuals access the Oracle ($O$); alternatively, divergent opinion sets may persist, if individuals only include $O$'s opinion if it falls within their bounded confidence envelopes (BCEs).

**Conjecture 2**: If all individuals have access to $T$, then all opinions will converge to $T$; access to $O$ will, all other parameters being equal, decrease the time taken to convergence (again, depending on whether $O$'s opinions are always accessed, or only accessed within the BCEs).

**Conjecture 3**: If only some individuals have access to $T$, then opinions may or may not converge to $\tau$; convergence to $\tau$ or, alternatively, a reduction in polarization will tend to occur if some or all individuals have access to $O$.

**Conjecture 4**: If only some individuals have access to $T$ and $O$, then a reduction in polarization will tend to occur if $O$'s opinion comes only from those individuals with access to $T$, rather than the entire community.

In the next section, I describe the new HK model with an AI extension, HKAI.

**The HKAI and its Variants**

Consider a community of $n$ individuals. The $i^{\text{th}}$ individual has opinion $x_i(t)$ at time $t$. At a following time $t+1$, the opinion of each individual is updated as:

$$x_i(t+1) = \alpha_i \tau + (1-\alpha_i)[w'_{i1} x_1(t) + w'_{i2} x_2(t) + \cdots \qquad (1)$$
$$+ w'_{i(i-1)} x_{i-1}(t) + w'_{ii} x_i(t) + w'_{i(i+1)} x_{i+1}(t)$$
$$+ \cdots + w'_{i(n-1)} x_1(t) + w'_{in} x_n(t)]$$

where $\tau$ is the true value of the opinion; $\alpha_i \in [0,1]$ is the relative weight that individual $i$ assigns to any evidentiary support that they might have of $\tau$; $(1-\alpha_i)$ is the weight that $i$ assigns to the opinions of others; and $w_{ij}$ is the weight that individual $i$ assigns to individual $j$'s opinion; when $w_{ij} = 0$, individual $i$ completely disregards the opinion of individual $j$. Hegselmann and Krause suggest that weights are chosen such that for each $i$,

$$\sum_j w_{ij} = 1$$

With the BCE extension to the HK model, for individual $i$,

$$w'_{ij} = \begin{cases} k_i w_{ij} & |x_i(t) - x_j(t)| \leq \varepsilon \\ 0 & \text{otherwise} \end{cases}$$

Here, $\varepsilon$ is a threshold above which individual $i$ ignores the opinion of individual $j$. For simplicity, I set $w_{ij} = 1$, for all $i$ and $j$; and $k_i$ as the reciprocal of the number of individuals within the BCE of $i$ (this has the simple effect of averaging the opinions of all individuals within $i$'s BCE). It is important to note here that $i$'s own opinion is also contained within $BCE_i$ and is not given any special weight (other authors have explored how giving more weight to an individual's own opinion might change the dynamic, e.g., (Khateri, Pourgholi et al. 2019, Wang, Rong et al. 2023)).

For further computational ease, in our implementation of the HK model, I have set

$$\alpha_i = \begin{cases} \alpha & \text{if } i \text{ has access to } \tau \\ 0 & \text{otherwise} \end{cases} \qquad (1b)$$

Thus, we can rewrite (1) as:

$$x_i(t+1) = \alpha_i \tau + (1-\alpha_i) k_i \sum_{j=1}^{n} x_j(t) \qquad (2)$$

We are now ready to introduce AI Oracles into the model; thus, we modify (2) as follows:

$$x_i(t+1) = \beta_i \frac{1}{|AI_i(t)|} \sum_{m \in AI_i(t)} O_m(t) + (1-\beta_i)[\alpha_i \tau \qquad (3)$$
$$+ (1-\alpha_i)k_i \sum_{j=1}^{n} x_j(t)]$$

There are a few important points to note with (3):
1. In our model, $AI_i(t) = \{O_1(t), \ldots, O_{r(i)}(t)\}$ is the set of Oracles that $i$ can access, and there is no restriction on the number of Oracles that can be included in HKAI; however, since our goal in this paper is to introduce the concept of an Oracle and describe its basic dynamics, I set $r(i) = 1$. Henceforth, $O(t)$ will refer to the opinion of this single Oracle at time $t$. As I will shortly explain, the value of the Oracle's opinion changes over time, hence the indexation by $t$.
2. At each time step, $i$ may choose to always access the Oracle's opinion; alternatively, $i$ may choose to access the Oracle only if $O(t-1)$ falls within $i$'s BCE at time $t$. This is an option that is set at the start of each simulation.
3. Third, as we did with $\alpha_i$, we set

$$\beta_i = \begin{cases} \beta & \text{if } i \text{ accesses } O \\ 0 & \text{otherwise} \end{cases} \qquad 3b)$$

There are two ways that the value of $O(t)$ is obtained: the Oracle may access the opinions of all individuals in the community, or only the opinions of individuals in the community who have access to the true value, $\tau$. Thus, for all individuals in $Oset$, the collection of individual opinions that the Oracle accesses,

$$O(t) = \frac{1}{|Oset|} \sum_{i \in Oset} x_i(t-1)$$

that is, $O(t)$ is the average opinion of individuals within $Oset$ at time $t$.

In any community, it is possible that only a fraction of individuals, $P$, has access to $\tau$. Similarly, only a fraction of individuals, $Q$, may have access to the Oracle. It is easy enough to implement partial access to either $\tau$ or $O$ by modifying (1b) and (4b) so that access for each $i$ depends on two independent Bernoulli random variables with parameters $P$ and $Q$, respectively, in addition to other criteria that may also determine access (specifically, whether $i$ chooses to include the Oracle regardless of whether it falls within $i$'s BCE, or not).

**Parameters, Simulations, and Summary Statistics**

All simulations were performed using R version 4.4.1 ((Team 2024) in the RStudio IDE (Team 2020), with code developed with the assistance of ChatGPT-4o (OpenAI 2024). Simulations were designed to test each of the 4 conjectures listed above. The number of individuals in each community was set to 500. Initial opinion values were sampled from a uniform random distribution bounded between 0 and 1 ( thus, with a mean of 0.5). The true value, $\tau$, was set to 0.8 in all simulations. The following parameters were available in all simulations, with some functioning as switches, turning features on or off.

- *BCE threshold*, $\varepsilon$: in earlier simulations, $\varepsilon$ was set to 0.02, 0.04, 0.08, and 0.16. As I will indicate, in later simulations, $\varepsilon$ was fixed at 0.08, a value chosen because it delivered outcomes in the Goldilocks Zone, i.e., dynamical behaviors that did not exclusively lead to either polarization or convergence.
- *p_truth_access*, $P$: the probability that an individual would have access to $\tau$. If $P = 1$, then all individuals has access to $T$; if $P = 0$, no individual has access to $T$; otherwise $P$ took values of 0.0, 0.05, 0.2, 0.5, and 1.0.
- *p_oracle_access*, $Q$: the probability that an individual would have access to $O$. If $Q = 1$, then all individuals had access to $O$; if $Q = 0$, no individual had access to $O$; otherwise $Q$ took values of 0.0, 0.05, 0.2, 0.5, and 1.0.
- *truth_weight* $\alpha$ given to $\tau$ by each individual: set to 0.01, 0.02, 0.04, 0.08, 0.16, and 0.32. When $P = 0$, $\alpha$ is disregarded.
- *oracle_weight* $\beta$ given to $O$ by each individual: set to 0.01, 0.02, 0.04, 0.08, 0.16, and 0.32. When $Q = 0$, $\beta$ is disregarded.
- *Oracle strategy*: When only some fraction, $P$, of individuals have access to $T$, this parameter sets the strategy than an Oracle uses to collect (and average) opinions. This parameter can be set to "random" or "truth-biased". With "random", the Oracle samples the opinions of all individuals; conversely, with "truth-biased", the Oracle only samples the opinions of individuals with access to $\tau$. This parameter only applies when we test Conjectures 3 and 4. (The sampling strategy labelled "random" future-proofs simulations to allow the Oracle to sample only a random subset of opinions, rather than all opinions; however, in the results reported here, subset sampling was not applied).
- *Individual bias, BCEYN (for BCE, Yes or No):* This parameter sets the behavior of individuals with respect to how they access the Oracle. If BCEYN = 0, then individuals access the Oracle with weight $\beta$ regardless of whether the Oracle's opinion falls within their BCE. In contrast, if BCEYN = 1, then individuals access the Oracle's opinion only if it falls within their BCEs. Again, $Q = 0$ denies Oracle access to all individuals.

Five simulations were run for each combination of parameter values. In addition to plots of opinion trajectories, results were summarized by calculating the following statistics:

- *Convergence time*: the timestep, $c$, when, for all $i$, $|x_i(c) - x_i(c-1)| < 1 \times 10^{-4}$.
- *Polarization:* the sample standard deviation, $s$, of all opinions in the community, at time $c$:
  $s = \sqrt{\sum(x_i - \bar{x})^2/(n-1)}$
  where $\bar{x}$ is the mean of all opinions in the community.
- *Truth alignment:* the average of absolute differences of all opinions to the truth at time $c$:
  $\sum abs[x_i(c) - \tau]/n$.

To allow plots of opinion trajectories to include some time after convergence, each simulation was allowed to run for an additional 10 timesteps beyond the convergence time, before it was halted. Maximum number of time-steps for each simulation was set at 10,000, although convergence was always attained before this ceiling.

Opinion trajectory plots were produced in R. All other graphs were produced using JMP version 17.0.0 (Inc. 2025).

# Results

In this section I report on the outcome of the simulations, as these relate to Conjectures 1-4 above.

<u>Opinion dynamics with individual access to an Oracle, but without individual access to $T$</u>

**Conjecture 1.** In the absence of access to T, communities will converge to a common opinion if all individuals access the Oracle ($O$); alternatively, divergent opinion sets may persist, if individuals only include $O$'s opinion if it falls within their bounded confidence envelopes (BCEs).

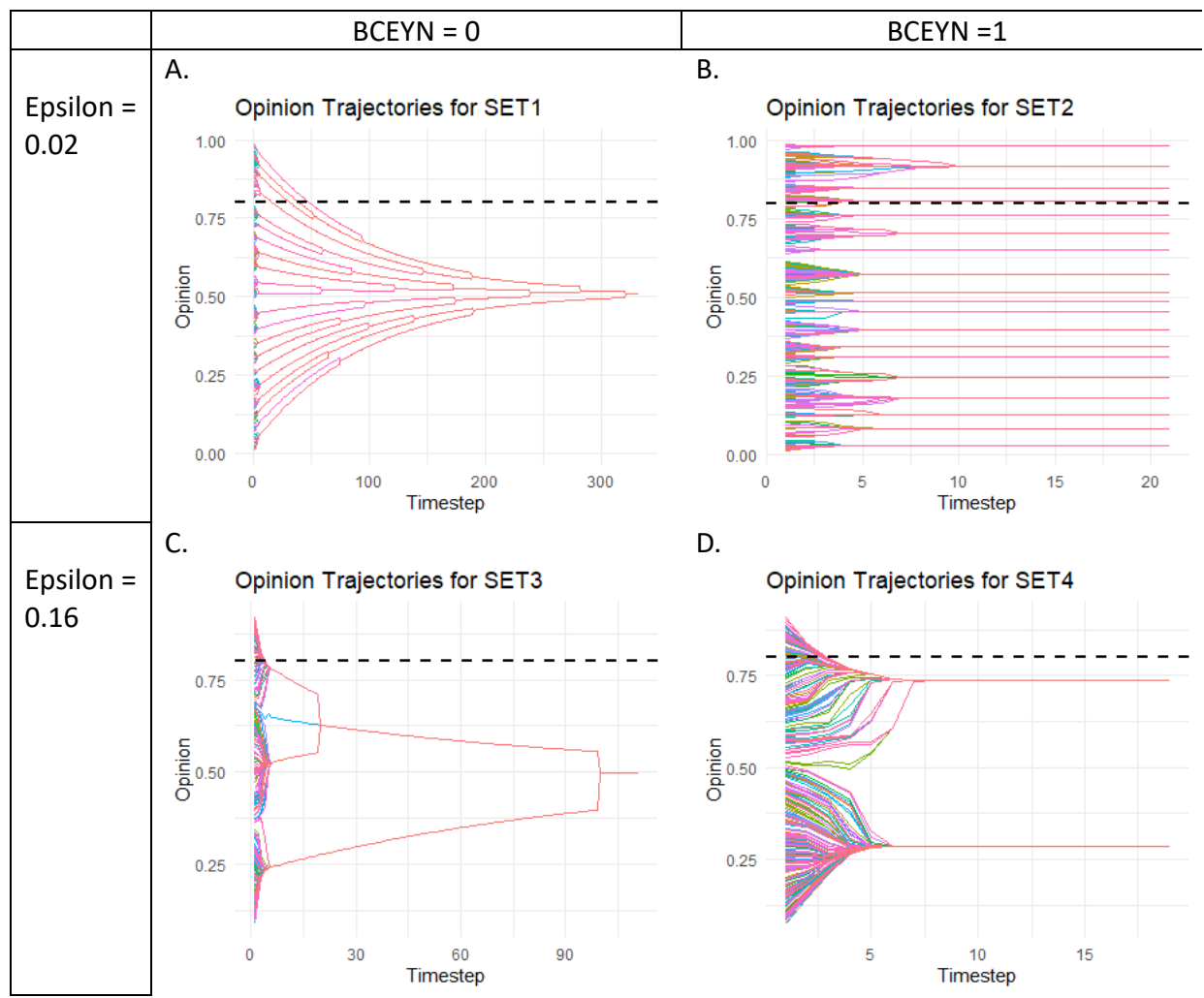

**Fig. 1.** Opinion trajectories, when individuals have access to an Oracle but no access to the truth, $T$. In all trajectories, individuals access the Oracle with weight $\beta = 0.01$. Columns and rows show the combination of $\varepsilon$ and BCEYN parameter values.

This conjecture is borne out by the results of our simulations. As can be seen in the sample of simulated opinion trajectories in Figures 1A and 1C, when BCEYN=0, opinion values converge to 0.5 (the mean of the uniform distribution for the initial set of opinion values). Conversely, when individuals only include the Oracle's opinion when it is within their BCEs, opinion trajectories cluster into different groups (Figures 1B and 1D).

Figure 2 summarizes how the accessibility ($Q = 1$) or inaccessibility ($Q = 0$) of the Oracle influences opinion polarization, as a function of the Oracle weights, the different values of the BCE threshold, e, and whether or not individuals always access the Oracle (BCEYN = 0) or access the Oracle only when its opinion falls within their BCEs. Polarization always decreases as e increases: this is to be expected, because high values of e mean that individuals accommodate more opinions within the BCE, and are therefore are more likely to converge. What is significant is the fact that when all individuals have access to the Oracle ($Q = 1$) and all individuals give some weight to the Oracle's opinions (BCEYN=0), then all opinions converge to a common value, regardless of the weight assigned to the Oracle's opinion.

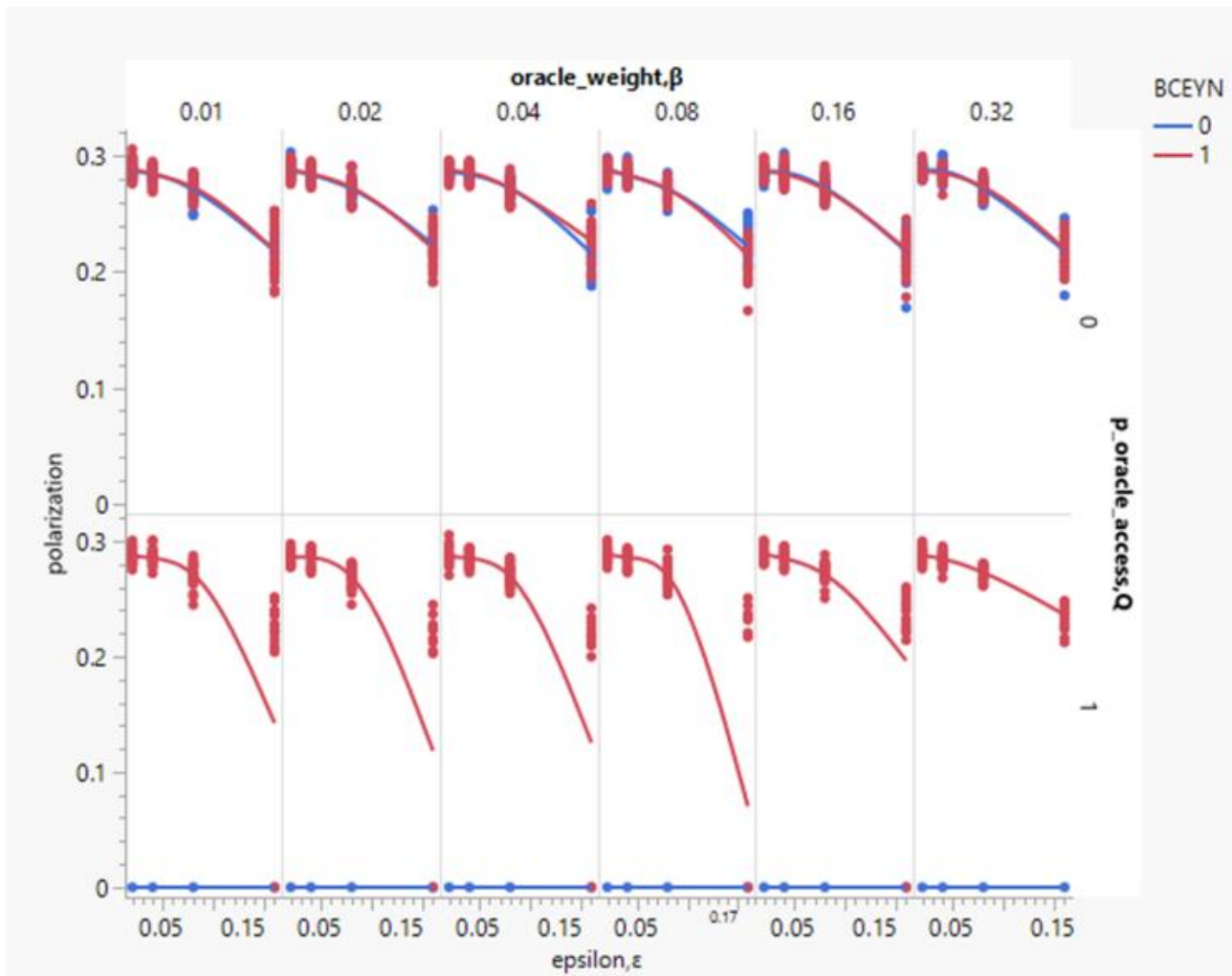

**Fig. 2.** Plot of polarization against $\varepsilon$, partitioned by the probability of access to $O$ ($Q = 0$ or 1), and $\beta$. Individuals do not have access to T. Note that when $Q = 0$, $\beta$ has no influence on opinion values, and the only variable impacting polarization is $\varepsilon$. In contrast, when $Q = 1$, there is no polarization when BCEYN = 0 (blue line). When $Q = 1$ and BCEYN=1, some simulations, but not all, also lead to complete convergence (i.e., no polarization) of opinions. (Note that the Oracle sampling strategy has no effect on the Oracle's opinion values in these simulations).

In contrast, when individuals only accommodate the Oracle's opinion when it falls within their BCEs (i.e., BCEYN = 1), polarization shows the same qualitative pattern as when an Oracle is inaccessible, decreasing as $\varepsilon$ increases. Nonetheless, there are some instances where there is a precipitous drop in polarization as $\varepsilon$ increases. This happens when the weight assigned to the Oracle is low and $\varepsilon$ is high, suggesting that under these conditions the Oracle may influence individual opinions gradually, at a pace that allows more opinions to be co-opted into larger and fewer groups (e.g., see Fig 1D), until occasionally all that remains is a single cluster of opinions.

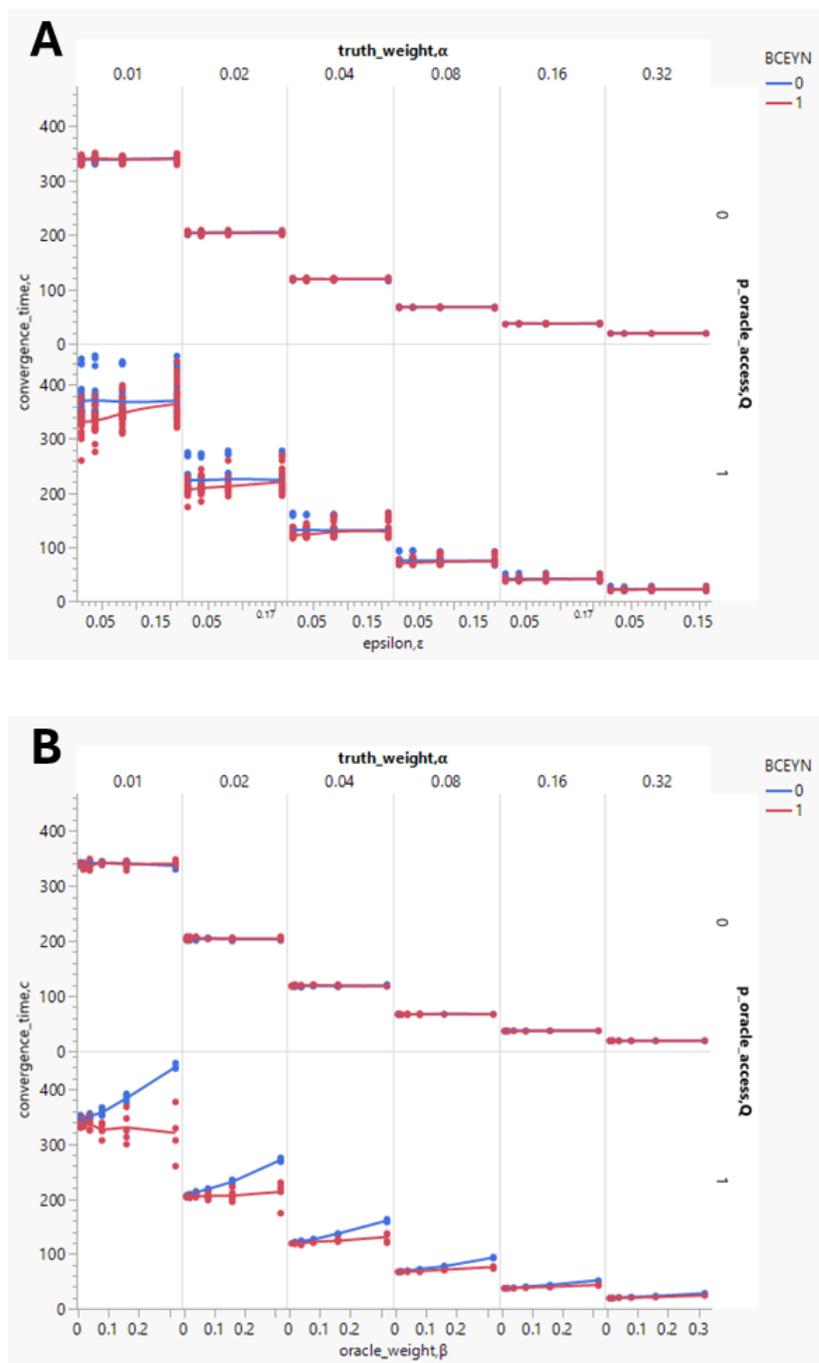

**Fig. 3.** Plots of convergence time against (A) $\varepsilon$, or (B) $\beta$, with and without Oracle access, when all individuals have access to T. For all simulations, $\varepsilon = 0.02$. When $Q = 0$, patterns of convergence times are identical in both A and B, as expected: these times vary with $\alpha$. In contrast, when Oracle access is available, universal access (i.e., BCEYN = 0) can lead to an increase in convergence times, particularly when $\alpha$ is low. (Note that the Oracle sampling strategy has no effect on the Oracle's opinion values in these simulations).

Opinion dynamics when all individuals have access to $T$

**Conjecture 2.** If all individuals have access to $T$, then all opinions will converge to $T = \tau$; access to $O$ will, all other parameters being equal, decrease the time taken to convergence (again, depending on whether $O$'s opinions are always accessed, or only accessed within the BCEs).

Here, I encounter the first set of results that run counter to expectations. The first part of Conjecture 2 is true, and confirms what Hegselmann and Krause (Hegselmann and Krause 2006) found: when all individuals have access to $T$, no matter what weight is assigned to this access, opinions always converge to $\tau$ (data not shown). The convergence time to $T = \tau$, either with or without $O$, is a negative exponential function of $\alpha$, the weight assigned to $T$. Convergence time for any given value of $\alpha$ does not vary with $\varepsilon$.

However, our simulations also show that the second part of Conjecture 2 – that access to $O$ will decrease the time to convergence to $\tau$ – is not true; in fact, whether individuals choose to always access $O$ (BCEYN = 0), or access $O$ only when $O$'s opinion falls within their BCE (BCEYN = 1), Figure 3 indicates that access to $O$ will have a significant effect on convergence time. This is most apparent when BCEYN = 0, $\alpha$ is low, and the weight $\beta$ assigned to $O$ is high. In these instances, individuals first converge to the common opinion value of the Oracle, before gradually moving towards $\tau$ (Figure 4A). In contrast, when individuals only include the Oracle's opinion when it falls within their BCE, average convergence time is similar to that seen when there is no access to $O$ (Fig 3 and F4B).

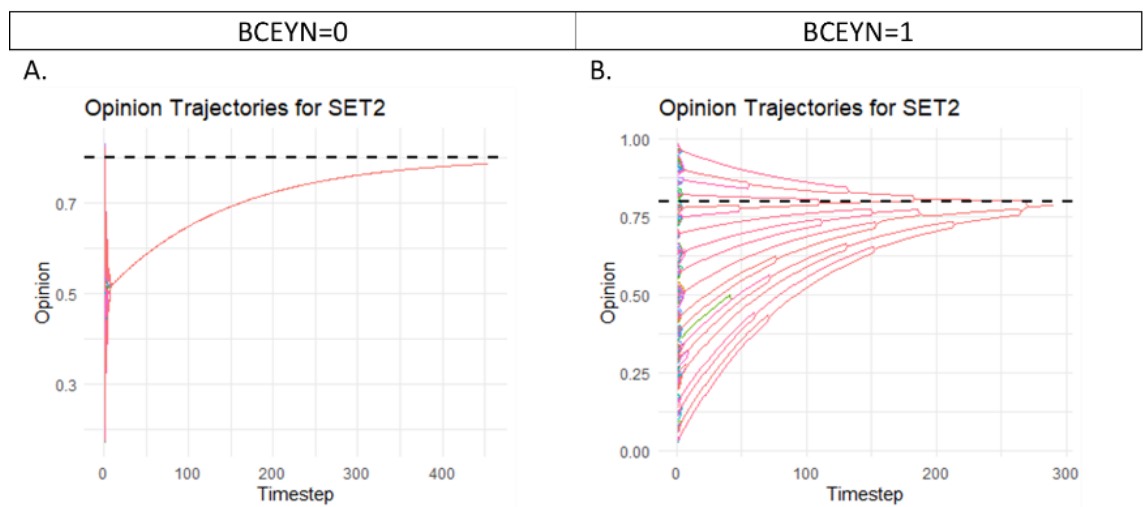

**Fig. 4.** Two opinion trajectory plots sampled from simulations shown in Fig. 3, when (A) BCEYN = 0 and (B) when BCEYN = 1. In these simulations, $\alpha = 0.01$ and $\beta = 0.32$. In (A), individual opinions rapidly converge to the Oracle's opinion value, before slowly converging on $T = \tau$. In (B), individual opinions move relatively quickly to converge on $T = \tau$.

The simulations also show that an Oracle increases the variance of convergence times, particularly at low values of $\alpha$. Thus, although an Oracle does not prevent all individuals converging on $\tau$ when everyone has access to $T$, it can increase the length of time it takes for this to happen.

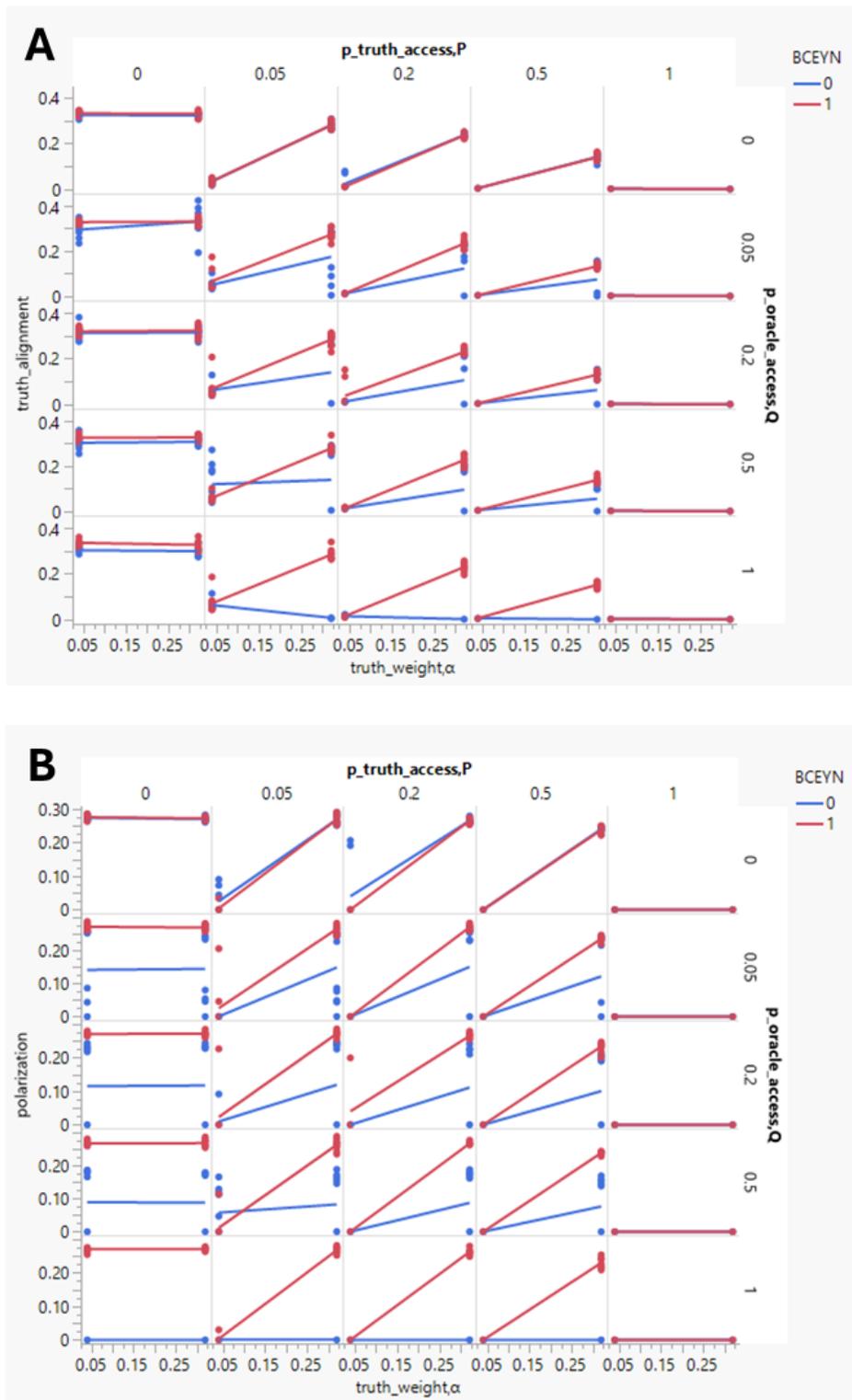

**Fig. 5.** Plots of (A) truth alignment or (B) polarization against $\alpha$, partitioned by different values of $P$ and $Q$. For all simulations, $\varepsilon = 0.08$, the Oracle sampling strategy is random, and $\alpha$ and $\beta$ only take values of either 0.04 or 0.32. (A) When there is no access to T, opinions are not closely aligned to $\tau$, but as more individuals have access to $T$ (i.e., as $P$ increases), opinions become more aligned to $\tau$. Universal access to the Oracle (BCEYN = 0), tends to increase this alignment. (B) If all individuals have access to O, then universal access ensures that there is no polarization.

Opinion dynamics when only a proportion of individuals have access to $T$ and $O$

**Conjecture 3.** If only some individuals have access to $T$, then opinions may or may not converge to $\tau$; convergence to $\tau$ or, alternatively, a reduction in polarization will tend to occur if some or all individuals have access to $O$.

Hegselmann and Krause (Hegselmann and Krause 2006) showed that when only some individuals have access to $T$, then there is no guarantee that all opinions will converge to $\tau$. Our simulations bear this out (Fig 5A). In fact, when there is no Oracle access (i.e., $Q = 0$), then as truth weight $\alpha$ increases, opinions are more polarized and the average opinion moves further from the true value (see Fig 6A and 6C). As Hegselmann and Krause (Hegselmann and Krause 2006) note, this happens because when only some individuals have access to $T$, and weight their truth access very highly, their opinions coalesce too rapidly, leaving behind others in the community with sufficiently different opinion values.

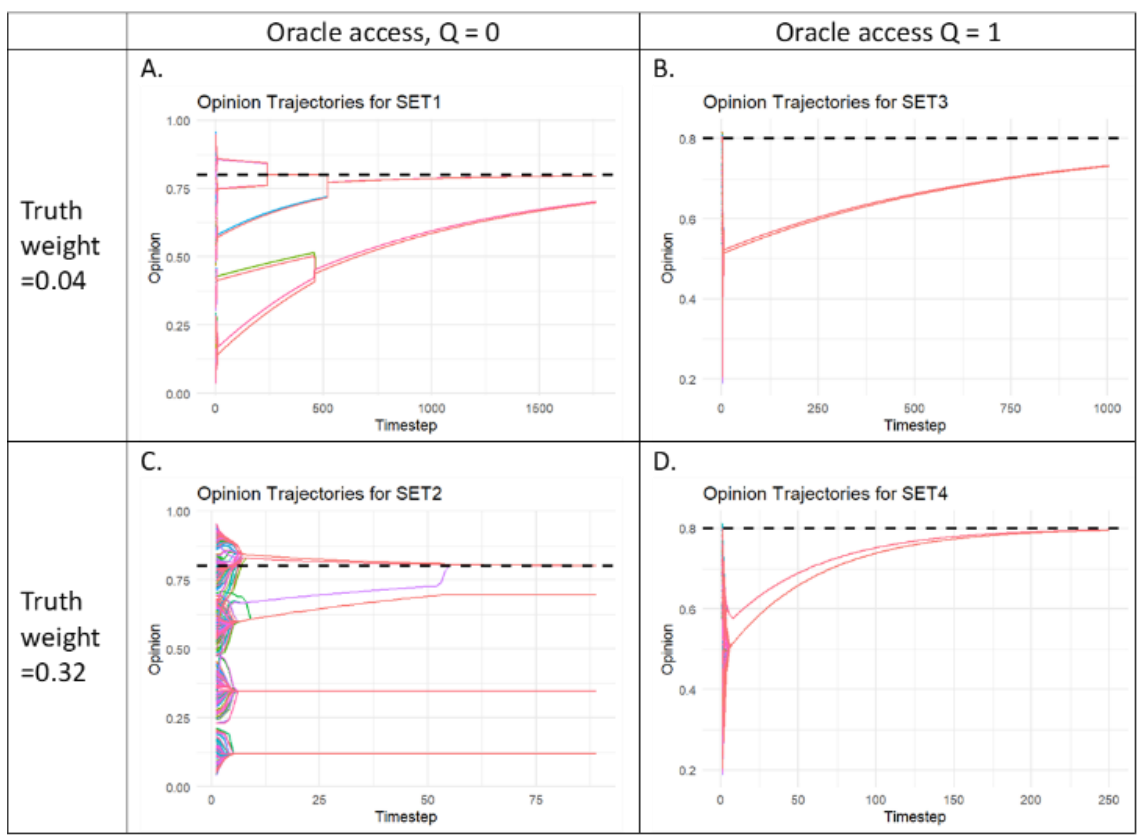

**Fig. 6.** Sample opinion trajectories from simulations in Fig. 5, under different values of $\alpha$ and Oracle access, $Q$. When $Q = 1$, then BCEYN = 0, in these simulations. Oracle sampling strategy is random. Importantly, when $\alpha$ is high, and individuals do not have access to an Oracle, opinions polarize (C). In contrast, when $\alpha$ high, but an Oracle is available and universally accessed (i.e., BCEYN = 0), then opinions converge on $\tau$.

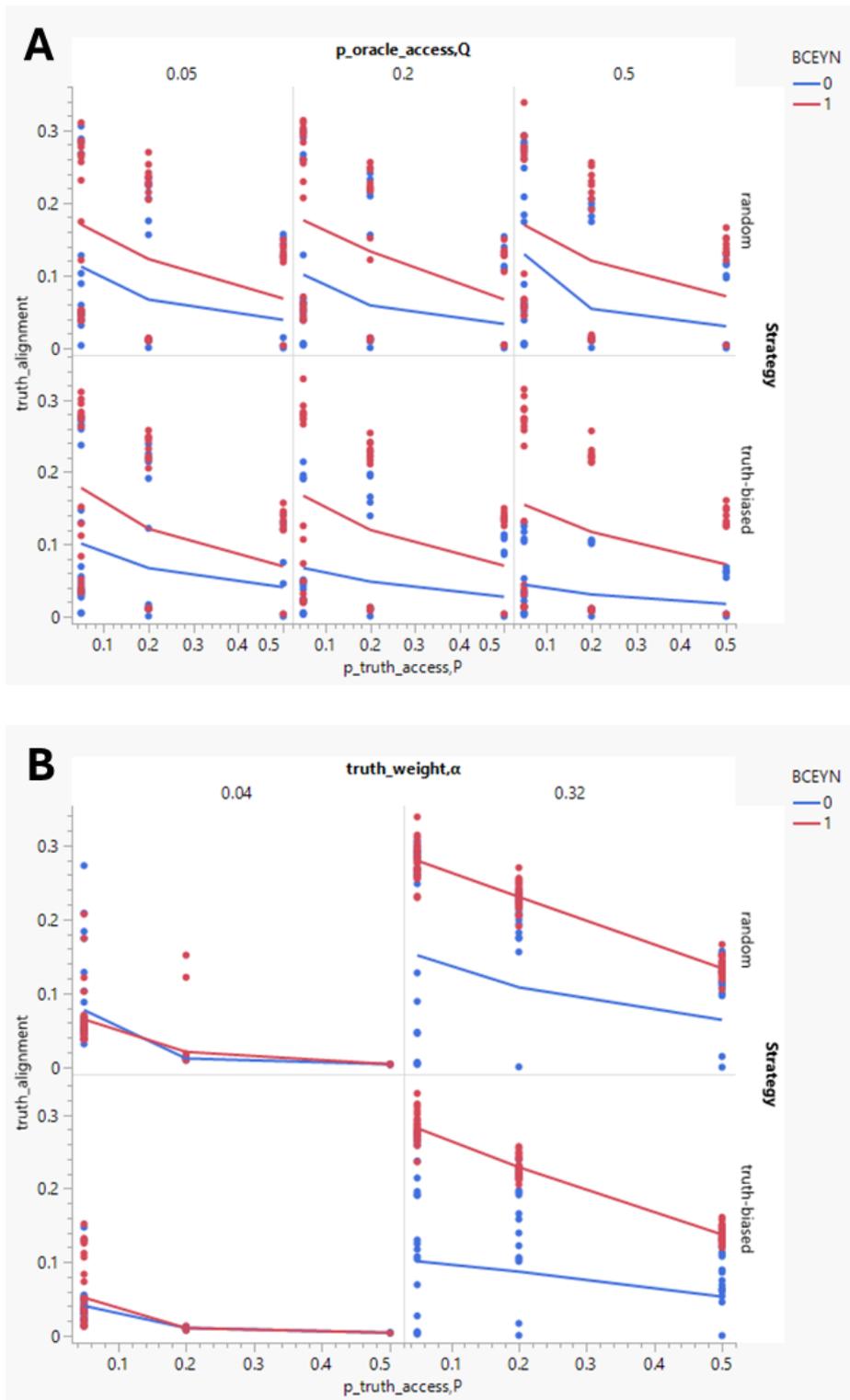

**Fig. 7.** Plots of truth alignment against $P$, partitioned by Oracle sampling strategy and (A) $Q$ and (B) $\alpha$. For all simulations, $\varepsilon = 0.08$, $P > 0$, and $Q > 0$, and $\alpha$ and $\beta$ only take values of either 0.04 or 0.32. Opinion alignment to $T$ depends heavily on $\alpha$, the number of individuals, $P$, with access to $T$, and whether individuals access the Oracle universally. Differences in $Q$ and Oracle sampling strategy show no substantial effect on truth alignment.

When an Oracle is introduced into the simulation, the situation changes, particularly when individuals access $O$ (i.e., BCEYN=0). Under these conditions, even if only a small fraction of individuals have access to $O$, average opinions tend to end up closer to $\tau$ (Fig 5A) and polarization decreases (Fig 5B), as truth weights increase. In other words, an Oracle dampens the effect of high truth weights and the consequent rapid coalescence of opinions amongst those who have access to the truth, so that even individuals without truth access are taken along. Examples of the opinion trajectories under different values of truth weights, and the accessibility or inaccessibility of the Oracle are shown in Figure 6. For instance, if only 5% of individuals have access to the truth, but all individuals have access to the Oracle, all opinions will converge to $T$ (Fig 6D) if sufficiently high weight is assigned to the Oracle's opinion.

Training the Oracle on individuals with access to $T$

**Conjecture 4.** If only some individuals have access to $T$ and O, then a reduction in polarization will tend to occur if O's opinion comes only from those individuals with access to $T$, rather than the entire community.

As noted above, researchers have tried to "cure" generative AI's propensity for hallucinations by training LLMs on domain specific knowledge. In this case, an AI Oracle will likely synthesize opinions from those individuals that have access to $T$ (equivalent, for example, to restricting the pre-training corpus only to academic articles or research databases).

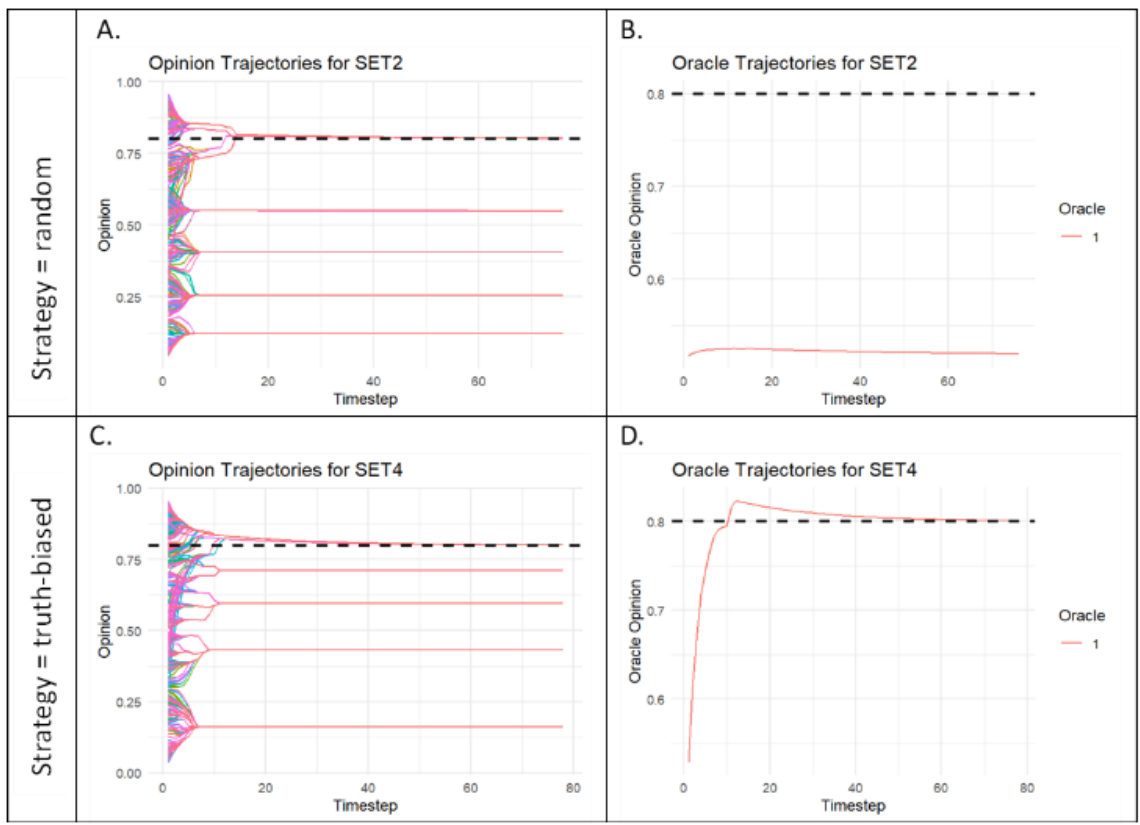

**Fig. 8.** Example opinion (A,C) and Oracle (B, D) trajectories, under different Oracle sampling strategies. For both simulations, BCEYN=1, $\varepsilon = 0.08$, $P = Q = 0.05$, and $\alpha = \beta = 0.05$. Note that even as both simulations show strong polarization, when the Oracle sampling strategy is truth-biased, the Oracle's opinion converges on $T = \tau$.

In our simulations, I tested this by applying two Oracle sampling strategies: "random" in which an Oracle samples all individuals regardless of whether they have access to $T$; or "truth-biased", in which an Oracle's opinion is formed by averaging the opinions of only those individuals with access to $T$.

Contrary to our conjecture, neither sampling strategy appeared to strongly influence the average opinion alignment with the truth, $\tau$ (Fig. 7A). Instead, as I have shown before, the most significant contributors to truth alignment (Fig. 7B) is truth weight (with lower values leading to greater alignment), and whether all individuals access the Oracle (where greater alignment is obtained if all individuals who have access to the Oracle, use the Oracle, i.e., BCEYN = 0).

It is important to note that our simulations speak only to the effect of domain-specific training on the alignment of individuals' opinions to the truth. They do not speak to the improvement or accuracy of the AI Oracle's "opinion" when it is trained exclusively on individuals with domain knowledge. In our simulations, truth-biased Oracles always converge on the true value, t, whereas Oracles that access all opinions may not, particularly when BCEYN= 1 (see Figure 8 for examples of opinion and corresponding Oracle trajectories).

**Discussion**

In this paper, I introduce a simple extension to the Hegselmann-Krause Bounded Confidence Model to incorporate an AI Oracle. The Oracle's main task is to synthesize the opinions from all, or a subset, of the community. In the model, the Oracle does this either by averaging the opinions of all individuals at each time step, or only those individuals who also have access to the true value, $T$. The latter case is equivalent to training the Oracle on individuals with domain-specific knowledge, i.e., some knowledge of $T$.

The oft-quoted adage attributed to George Box (Box 1976) that "all models are wrong but some models are useful" is particularly apt for our model. It is likely that the very simple characterization of how an AI Oracle operates is wrong precisely because it is too simple, but I argue that as a baseline model it is useful, because it generates patterns that serve as testable hypotheses:

1. If individuals only have access to the Oracle, without any access to $T$, then all opinions will converge on a common value if all individuals access the Oracle.
2. If all individuals also have access to $T$, then although all opinions will ultimately converge to $T$, if the Oracle is universally accessed, then time to convergence may be delayed.
3. If only some individuals have access to $T$, there is no guarantee that opinions will ultimately converge to $T$, but universal access to the Oracle will guarantee convergence to $T$, if individuals given the Oracle's opinion sufficiently high weights.
4. If only some individuals have access to $T$ and $O$, it makes no difference whether the Oracle samples opinions only from those with access to $T$ or from the entire community: in both cases, universal access to $O$ decreases the average distance of individual opinions from $T$, at approximately equivalent rates. Thus, although the Oracle's opinion value converges to $T$ when it only accesses individuals who have some knowledge of $T$, this makes little difference to how individuals opinions differ from $T$.

Each of the statements (1) – (4) are testable, insofar as it is possible to imagine appropriately designed social science research that addresses each hypothesis. In fact, simple models such as the one developed here are most interesting (and most useful) when empirical data conflict with what is observed. When this happens, it provides an opportunity to refine the modelling assumptions, or extend the model by adding new parameters. By iterating through hypothesis refinement and experimental testing, we expect to get closer to *T*!

Even in the absence of empirical tests, there are obvious extensions to the HKAI model that are reasonable. For instance, it is already possible to introduce two or more Oracles, and it would be interesting to see how having more (potentially, competing) Oracles influences opinion trajectories. It is also reasonably straightforward to introduce noise into the Oracle's opinion to simulate the propensity for current gAIs to "hallucinate". In fact, Su et al (Su, Chen et al. 2017) proposed something similar when they introduced noise to a leader agent's fixed opinion, and showed that over time, all other opinions synchronized with that of the leader. Will this happen with Oracles?

Finally, the idea that communities of individuals can converge on an answer that is close to the Truth, as exemplified by the story of Galton's ox, has been termed "the wisdom of crowds". This is often contrasted with "groupthink" whereby groups of individuals may (self-)suppress dissent, and potentially reinforce biases, so as to achieve consensus (Solomon 2006); when there are many such groups, polarization may eventuate. The literature on both of these social phenomena is large, and this is not the place to go into any detail except to say that the patterns we observe with the HK and HKAI models lend themselves to quantifying "the wisdom of crowds" (as truth alignment) and groupthink (as polarization) as additive and orthogonal sources of variation of the total variation of opinion values. I have not done this here, but it remains an active research goal.

## Acknowledgements

I am grateful to Andrew Withy of the Philosophy Department at The University of Auckland for his comments during the conception of this work, and on drafts of this manuscript.

## Code availability

The R script for the HKAI model is available at: https://github.com/allengrodrigo/HKAI.